\documentclass[12pt]{article}
\usepackage{epsfig}
\usepackage{float}
\usepackage{graphicx}
\begin{document}
\rightline{NKU-2015-SF4}
\bigskip

\newcommand{\be}{\begin{equation}}
\newcommand{\ee}{\end{equation}}
\newcommand{\noi}{\noindent}
\newcommand{\ra}{\rightarrow}
\newcommand{\bib}{\bibitem}
\newcommand{\refb}[1]{(\ref{#1})}

\newcommand{\bff}{\begin{figure}}
\newcommand{\eff}{\end{figure}}

\begin{center}
{\Large\bf Regular black holes in de Sitter universe: scalar field perturbations and quasinormal modes}

\end{center}
\hspace{0.4cm}
\begin{center}
Sharmanthie Fernando \footnote{fernando@nku.edu}\\
{\small\it Department of Physics \& Geology}\\
{\small\it Northern Kentucky University}\\
{\small\it Highland Heights}\\
{\small\it Kentucky 41099}\\
{\small\it U.S.A.}\\

\end{center}

\begin{center}
{\bf Abstract}
\end{center}
The purpose of this paper is to study quasinormal modes (QNM)  of a regular black hole with a cosmological constant due to scalar perturbations. A detailed study of the QNM frequencies for the massless scalar field were done by varying the parameters of the theory such as the mass, magnetic charge, cosmological constant, and the spherical harmonic index. We have employed the sixth order WKB approximation to compute the QNM frequencies.  We have also proved analytically that the $l=0$ mode for the massless field reach a constant value at late times. We have approximated the near-extreme regular-de Sitter black hole potential with the P$\ddot{o}$schl-Teller potential to obtain exact frequencies. The null   geodesics of the regular-de Sitter black hole is employed to describe the QNM frequencies at the eikonal limit ($ l >>1$).

\hspace{0.7cm}


\section{ Introduction}

It is a well known fact that our universe is expanding at an accelerated rate. There are many observational evidence for this  \cite{perl}\cite{riess}\cite{sper}\cite{teg}\cite{sel}. One of the simplest candidates to describe the mysterious dark energy causing the acceleration is the cosmological constant $\Lambda$ introduced by Einstein to General Relativity to describe a static universe.  The space-time with a positive cosmological constant is known as the de-Sitter space-time. Hence,  our universe may be described by de Sitter geometry. There are several reasons to study de Sitter spaces. In high energy physics, there is a holographic duality relating quantum gravity on the de Sitter space to conformal field theory on a sphere of one dimension lower \cite{witten} \cite{stro} known as dS/CFT correspondence. Examples along those lines include, Kerr-de Sitter space-time in various dimensions studied by Dehghani \cite{deh} and holographic description of the three dimensional Kerr-de Sitter space with gravitational Chern-Simons term  studied  by Park \cite{park}. One other reason for de Sitter space-times to gain lot of attention is the need to understand it in from the point  of string theory. String theory is  one of the popular candidates for a theory of quantum gravity. Hence, a fully satisfactory de Sitter solution to string theory has to be found. Towards this goal, classical solutions in Type IIB string theory have been studied by Dasguptha et.al \cite{das}. Another work studying classical de Sitter solutions in string  theory has been published by Danielsson et.al. \cite{dan}.

Usually, a black hole is a space-time with a horizon and a singularity covered by the horizon. Contrary to this notion, there are black holes where there is a horizon but without a singularity at $ r =0$. Such black holes are called ``regular'' black holes. The first regular black hole was constructed by Bardeen \cite{bardeen}  \cite{fernando7} and interpreted as a magnetic monopole by Ay$\acute{o}$n-Beato and Garc$\acute{i}$a \cite{gar1}. There are several other regular black hole solutions proposed by Ay$\acute{o}$n-Beato and Garc$\acute{i}$a \cite{gar2}\cite{gar3}\cite{gar4}\cite{gar5}. In this paper, we are interested in studying one of the most interesting regular black holes proposed by Ay$\acute{o}$n-Beato and Garc$\acute{i}$a \cite{gar2} and inteprited by Bronnikov \cite{bron}. This black hole is static, spherically symmetric and is described by two parameters, mass $M$ and magnetic charge $Q$. For $\frac{ r}{M} >>1$ and for small $\frac{|Q|}{M}$, such black holes have similar properties as the well known Reissner-Nordstrom solution. This family of black hole solutions are referred to as ABGB black holes. Here we study the ABGB black hole in a de Sitter universe  presented by Matyjasek et.al \cite{mat}. Since the current universe may be a one with a positive cosmological constant, it is important to understand regular black hole embedded in a de Sitter universe.

When a black hole is externally perturbed, the response can be described by three stages. The initial stage entirely depend on the initial data of the external perturbation. The intermediate stage correspond to damped oscillations with incomplete set of complex frequencies. These modes are called quasi-normal modes (QNM)  and depend only on the macroscopic properties of the black  hole such as the mass, spin and the charge. Measuring QNM frequencies using current gravitational wave detectors such as LISA, VIRGO and LIGO  give a direct approach to detect black holes \cite{ferra}. Other than for experimental reasons, there are also several theoretical aspects of black holes which are studied by using QNM frequencies.  Due to the famous AdS/CFT correspondence,  many works have been devoted to study QNM's of black holes in anti-de Sitter space \cite{vitor}\cite{horowitz}. Black hole quantum spectrum via QNM were presented by Corda in \cite{corda}:  the black hole model used by Corda is somewhat similar to the semi-classical Bohr's model for the hydrogen atom. Another reason to create interest on QNM's was the conjecture by Hod \cite{hod3} relating quantum properties of black holes and asymptotic values of frequencies of QNM's. There are many works devoted to   compute asymptotic QNM frequencies due to this reason.  Some recent work have addressed relation between QNM's and hidden conformal symmetry of wave equations on the black hole back ground  \cite{chen} \cite{kim1}.  Other than  black holes,   naked singularities also have been studied from the QNM point of view \cite{saa}. An excellent review on QNM's is written by Konoplya and Zhidenko \cite{kono1}.  Given all the  above, it is worthwhile to study QNM's of  regular black holes with nonlinear sources.

The paper is organized as follows: in section 2 an introduction to the regular-de Sitter black hole is given. In section 3 the scalar field perturbation is introduced. In section 4, the WKB approach to compute QNM are introduced.  Massless scalar perturbation with QNM frequencies are presented in section 5.  In section 6, analysis of $l=0$ mode of the massless scalar field is presented. In section 7, unstable null geodesics approach is used to compute QNM frequencies.  In section 8, P$\ddot{o}$schl-Teller method to compute QNM frequencies are presented. In section 9 the conclusion is  given.


\section{ Regular black hole with a cosmological constant (ABGB-de Sitter black hole)}

In this section we will present the basic characteristics of a regular black hole with a positive cosmological constant derived by Matyjasek et.al \cite{mat}. The action from which it was derived is,
\be \label{action}
S = \frac{ 1}{ 16 \pi G} \int \sqrt{ -g} \left[ ( R - 2 \Lambda) -  \mathcal{L}(F) \right]
\ee
Here $R$ is the scalar curvature and $\Lambda$ is the positive cosmological constant. $\mathcal{L(F)}$ is the Lagrangian for  non-electrodynamics component given by,
\be
\mathcal{L}(F) = F \left[ 1 - tanh^2 \left( \alpha \sqrt[4]{ \frac{Q^2 F}{2}} \right) \right]
\ee
Here $ \alpha = \frac{|Q|}{ 2 M}$ and $F = F_{\mu \nu} F^{ \mu \nu}$ with $F_{\mu \nu} =  \bigtriangledown_{\mu} A_{\nu} - \bigtriangledown_{\nu} A_{\mu}$. The only component of $F_{\mu \nu}$ which is non-zero for this solution is $F_{23} = F_{\theta \phi}$ given by,
\be
F_{23} = Q sin \theta
\ee
Here, $Q$ is the magnetic charge. Hence,
\be
F = \frac{ 2 Q^2}{ r^4}
\ee
Static spherically solutions for the action in eq.$\refb{action}$ with the given magnetic field was derived in \cite{mat} to be,
\begin{equation} \label{metric}
ds^2 = - f(r) dt^2 + \frac{ dr^2}{ f(r)} + r^2 d \Omega^2
\end{equation}
where,
\begin{equation}
f(r) = 1 - \frac{ 2 M} { r}\left(   1 - tanh \left( \frac{ Q^2}{ 2 M r} \right) \right) - \frac{ \Lambda r^2}{3}
\ee
This particular solution was named Ay$\acute{o}$n-Beato-Garc$\acute{i}$a-Bronnikov-de Sitter (ABGB-dS) solution.

The function $f(r)$ is finite at $r =0$.  $f(r)$ can have at most three distinct roots leading to three horizons:  $r_c$ is the cosmological horizon, $r_b$ is the event horizon and $r_i$ is the inner horizon. The region considered for the perturbations will be between $r_b$ and $r_c$. $f(r)$ also can have degenerate horizons: when $r_i = r_b$ they are referred to as  cold black holes and  when $r_b = r_c$, they are considered as charged Nariai black holes \cite{mat2}\cite{fernando3}.

When $Q=0$, the metric yeilds the Schwarzschild-de Sitter black hole and when $\Lambda =0$,  one obtain the ABGB black hole. Function $f(r)$  can be expanded for small $Q$ and for large $r$ as,
\be
f(r) \approx 1 - \frac{ 2 M}{r} + \frac{Q^2}{r^2} - \frac{\Lambda r^2}{3} - \frac{ Q^6}{ 12 M^2 r^4} + ...
\ee
Hence, it is clear that for large $r$ and for small $Q$, the line-element resembles the well known Reissner-Nordstrom-de Sitter black hole (RNdS) with the line element eq.$\refb{metric}$ with the function $f(r)$,
\be
f(r)_{RNdS} = 1 - \frac{ 2 M}{r} + \frac{Q^2}{r^2} - \frac{\Lambda r^2}{3}
\ee

The horizon structure of the ABGB-de Sitter black hole has been analyzed thoroughly in \cite{mat}. The regularized stress-energy tensor of the scalar, spinor and vector fields inside the degenerate ABGB-de Sitter black hole solution were studied by Matyjasek et.al \cite{mat2}.


\section{ Scalar field perturbations}

Having introduced the geometry of the black hole space-time, in this section we will present perturbations by a minimally coupled scalar field around this geometry. The equations of motion for such a field is given by the Klein-Gordon equation,
\be \label{klein}
(\bigtriangledown^2 - \mu^2) \eta = 0 
\ee
The scalar field $\phi$ can be decomposed into its partial modes as
\be
\eta(r, \theta, \phi,t) = \sum_{l,m}  e^{ - i \omega t}  Y_{ l, m} ( \theta, \phi) \frac{ R(r)}{ r}
\ee
Here, $Y_{l,m} (\theta, \phi)$ are spherical harmonics, and, $l$ and $m$ are the angular and the magnetic quantum numbers. $\omega$ is the frequency of the oscillations of the field. With the above decomposition, the radial component of the Klein-Gordon equation can be simplified to be,
\be \label{wave}
\frac{ d^2 R(r_*) }{ dr_*^2} + \left( \omega^2 - V_{scalar}(r_*)  \right) R(r_*) = 0
\ee
Here  $r_*$ is the tortoise coordinates which is given by,
\be \label{tor}
dr_* = \frac{ dr} { f(r)}
\ee
and  $V_{scalar}$ is given by,
\be \label{potscalar}
V_{scalar}(r) = f(r) \left( \frac{ l ( l + 1) } { r^2} + \frac{   f'(r)}{r} + \mu^2 \right)
\ee
The effective potential approaches zero at $ r = r_b$ ad $r = r_c$. It has a peak in between as demonstrated in Fig.$\refb{potmass}$.

To get approximations for $r_*$, one can expand $f(r)$ around $ r = r_{b,c}$ in a Taylor series as,
\be
f(r)  \approx f'(r_{b,c}) ( r - r_{b,c})
\ee
Since $ f'( r_c) < 0$, $r_*$ can be derived to be,
\be \label{torcos}
r_* \approx   - \frac{1}{ |f'(r_c)|} Log( r_c - r)
\ee
Hence, when $r \ra r_c$, $r_* \ra \infty$.
Similarly,  since  $ f'(r_b) >0$,  $r_*$ can be derived to be,
\be
r_* \approx \frac{1}{ |f'(r_b)|} Log( r - r_b)
\ee
Hence, when $r \ra r_b$, $r_* \ra - \infty$.
Hence the effective potential $V_{scalar} \ra 0$ when $ r_* \ra \pm \infty$.


\section{ Computation of QNM frequencies  of the scalar perturbations by WKB approach}

QNM for a black hole perturbed by a scalar field is given by the solutions of the wave eq.$\refb{wave}$. To find the QNM solutions one has to impose boundary conditions; it is purely ingoing field at the even horizon $r_b$ and purely outgoing fields at the cosmological horizon $r_c$.  The frequencies corresponding to the QNM are given by $ \omega = \omega_R + i \omega_I$; here $\omega_R$ is the oscillating  component  and $\omega_I$ is the damping component of the frequency. QNM for eq.$\refb{wave}$ with the above mentioned boundary conditions can be represented as,
\be
R(r_*) \ra exp( i \omega r_*);  \hspace{1 cm}  r_* \ra - \infty ( r \ra r_b)
\ee
\be
R(r_*) \ra exp( -i \omega r_*);  \hspace{1 cm}  r_* \ra + \infty ( r \ra r_c)
\ee
Since the effective potential has a peak and has the form of a potential barrier, to find the QNM frequencies, one can use the WKB approach.  WKB approach to find QNM frequencies of black holes was first developed by Iyer and Will for third order \cite{will}; it was   later extended to sixth order by Konoplya \cite{kono4}. The sixth order WKB approximation is employed to find QNM frequencies in several papers including, \cite{fernando7} \cite{gauss}  \cite{fernando8}.

In the sixth order WKB approach,   QNM frequencies are given by the expression,
\be
\omega^2 = - i \sqrt{ - 2 V''(r_{max})} \left( \sum^6_{i=2}  \Gamma_i  +   n + \frac{ 1}{2} \right) + V(r_{max})
\ee
Here, $r_{max}$ is  where $V(r)$ reach a  maximum and $V''(r)$ is the second derivative of the potential.  Expressions for $\Gamma_i$ can be found in \cite{kono4}. 



\section{ Massless scalar field perturbation}

First, we will focus on the massless scalar field perturbations where $\mu=0$. In this case, the effective potential $V_{scalar}$ depend on the parameters, $M, Q, l$ and $\Lambda$.
In Fig.$\refb{potmass}$, the potential is plotted by varying $M$. When the mass $M$ increases, the height of the potential decreases. In Fig.$\refb{potcharge}$,  $V_{scalar}$ is plotted by increasing the charge $Q$. When $Q$ is increased, the height of the potential increased. When the cosmological constant $\Lambda$ is increased, the height of the potential  decreases as given in Fig.$\refb{potlambda}$. In Fig$\refb{potl}$, the potential is plotted by varying $l$ where $l$ is positive. When $l$ increases,  the potential increases. In Fig$\refb{pot2}$, the potential is plotted for $l=1$ and $l=0$. It is interesting to notice that the potential for $l=0$ has a local minimum between the two horizons, $r_h$ and $r_c$ which is different from the behavior of the potential for other values of $l$.

\begin{figure} [H]
\begin{center}
\includegraphics{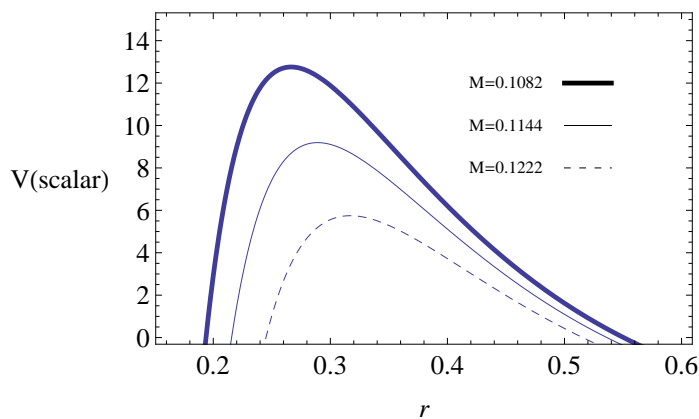}
\caption{The figure shows  $V_{scalar}(r)$ vs $r$. Here $Q = 0.086, \Lambda = 2.5$ and $l=2$}
\label{potmass}
 \end{center}
 \end{figure}
 
 \begin{figure} [H]
\begin{center}
\includegraphics{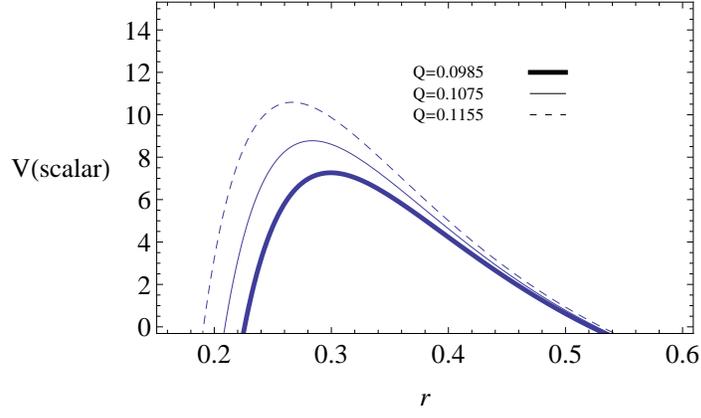}
\caption{The figure shows  $V_{scalar}(r)$ vs $r$. Here $M = 0.1222, \Lambda = 2.5$ and $l=2$}
\label{potcharge}
 \end{center}
 \end{figure}

\begin{figure} [H]
\begin{center}
\includegraphics{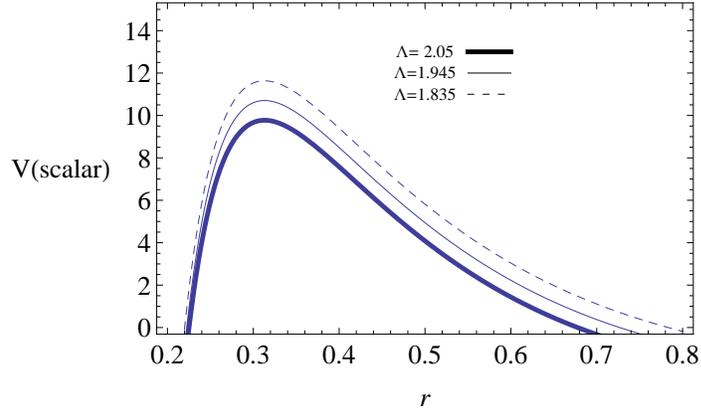}
\caption{The figure shows  $V_{scalar}(r)$ vs $r$. Here $Q = 0.1055, M = 0.129$ and $l=2$}
\label{potlambda}
 \end{center}
 \end{figure}
 
 \begin{figure} [H]
\begin{center}
\includegraphics{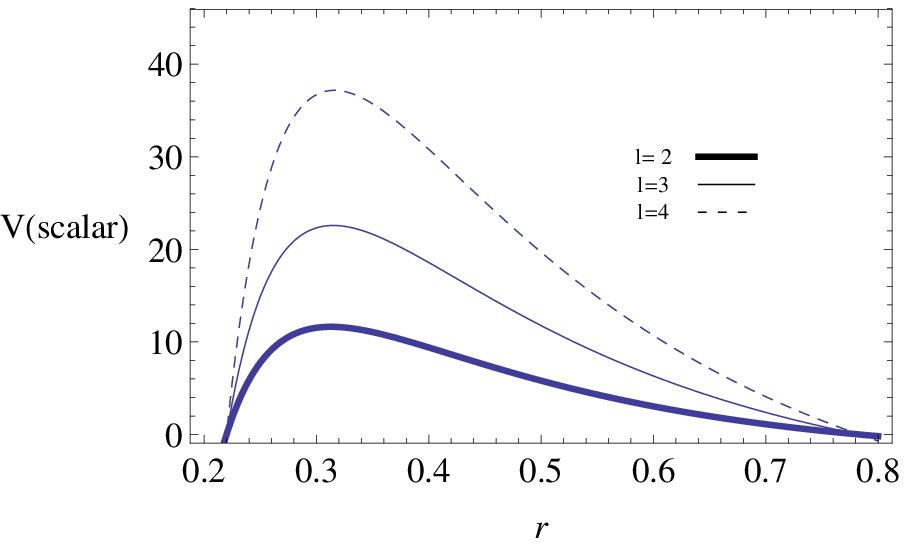}
\caption{The figure shows  $V_{scalar}(r)$ vs $r$. Here $Q = 0.1055, \Lambda = 1.835$ and $M=0.129$}
\label{potl}
 \end{center}
 \end{figure}
 
 \begin{figure} [H]
\begin{center}
\includegraphics{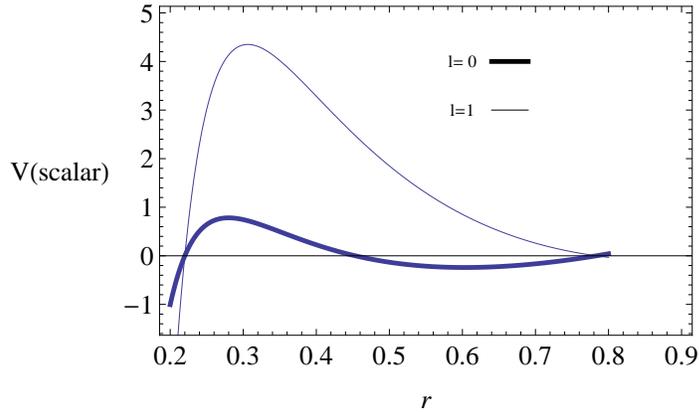}
\caption{The figure shows  $V_{scalar}(r)$ vs $r$ for $ l=1$ and $l=0$. Here $Q = 0.1055, \Lambda = 1.835$ and $M=0.129$}
\label{pot2}
 \end{center}
 \end{figure}


\subsection{Quasinormal modes of massless scalar perturbations for $ l >0$ modes}

In this section, we will discuss the properties of QNM frequencies for massless scalar fields for $ l>0$ modes. We want to mention that for all $\omega$ computed  $\omega_I$ was negative  leading to the conclusion that the black hole is stable under scalar perturbations for $l>0$. When plotting, we will only use the magnitude of $\omega_I$  in the rest of the paper.

First, we computed QNM frequencies by varying charge $Q$;  $\omega_R$ vs $Q$ and $\omega_I$ vs $Q$ are given in Fig.$\refb{omegacharge}$ (for $l=2$) and 
Fig.$\refb{omegacharge2}$ (for $l=1$). When $Q$ increases, $\omega_R$ increase. On the other hand when $Q$ increase, $\omega_I$ increase to a maximum and then start to fall off at large $Q$. Hence, there is a maximum value of $Q$ where the black hole has the most stability. We want to note that when $Q$ become large, the black hole is approaching the degenerate limit where $ r_i \approx r_b$.  Also, we have computed the ``Quality factor'' for the field which is defined as,
\be
Quality \hspace{0.2cm} Factor = \frac{ \omega_R}{ 2 |\omega_I|}
\ee
If the quality factor is large, then the field is considered a better oscillator. The quality factor for the scalar field is plotted in Fig.$\refb{qualitycharge}$ (for $l=2$) and 
Fig$\refb{qualitycharge2}$ (for $l=1$). The larger the charge, the larger the quality factor. Hence, the scalar fields are better oscillators for large charge.

We computed the radius of the black hole horizon $r_b$ for the same values of $Q$ and plotted $\omega$ vs $r_b$ in Fig.$\refb{omegaradius}$. First we want to mention that $r_b$ decreases when $Q$ is increased. Hence it is no surprise to  see that $\omega_R$ decreases when $r_b$ increases. On the other hand, the behavior of $\omega_I$ vs $r_b$ is similar to the behavior of $\omega_I$ vs $Q$: there is  a particular value of $r_b$ where $\omega_I$ reaches its maximum which is the most stable radius of horizon where stability is concerned.

For the same $Q$ values, we computed the Hawking temperature of the black hole; the plot of $\omega_I$ vs $T_b$ is given in Fig.$\refb{omegatemp}$.  In many black holes, $\omega_I$ seems to have a linear relation with the temperature \cite{horowitz}. In the regular black hole, there seems to be a strange relation between  $\omega_I$ and $T$; it is possible to have two values of $\omega_I$ for the same temperature.

Next, we computed $\omega$ by varying $\Lambda$. The plots for $\omega_R$ vs $\Lambda$ and $\omega_I$ vs $\Lambda$ are plotted in Fig.$\refb{omegacosmo}$. Both $\omega_R$ and $\omega_I$ decreases as $\Lambda$ increases.  Hence, a smaller $\Lambda$ is favored for a field decaying faster around the black hole. The quality factor as a function of $\Lambda$ is plotted in Fig.$\refb{qualitycosmo}$. One can see that the field is a better oscillator for small $\Lambda$.

QNM frequencies were also studied by varying $l$; $\omega_R$ vs $l$ is plotted in Fig.$\refb{sphereal}$ and $\omega_I$ vs $l$ is plotted in Fig.$\refb{spheima}$. When observed the behavior of $\omega_R$, it is clear that it varied linearly with $l$ for both $n=0$ and $n=1$.  When observed $\omega_I$,  it is clear that it decreases to reach a stable value when $l$ increases. for both $n=0$ and $n=1$. Hence, the field decays faster for small $l$ values.

We have also  plotted $\omega$ vs $n$ in Fig.$\refb{omegan}$. Since the accuracy of the WKB approach is greater for $ l > n$, we have chosen $ l =10$ for this computation. In this case, $\omega_R$ decreases with $n$ which is opposite to what we observe in a simple oscillating system with standing waves such as  a tight string.  On the other hand, $\omega_I$ increases linearly with $n$.

\begin{figure} [H]
\begin{center}
\includegraphics{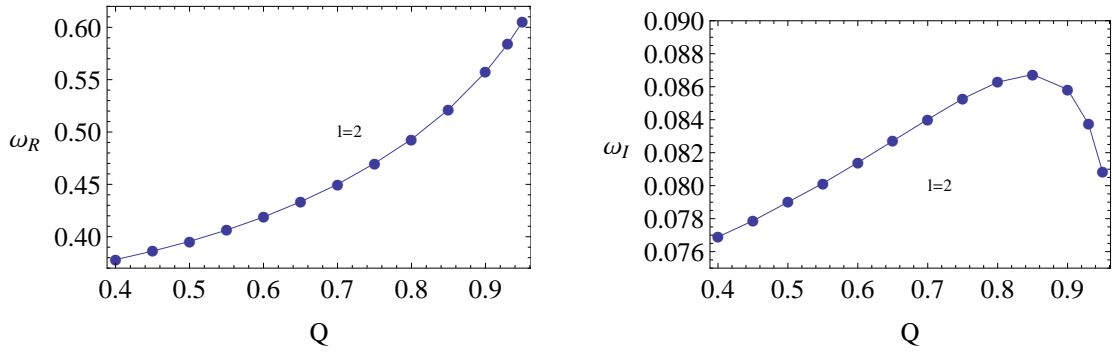}
\caption{The figure shows  $\omega$ vs $Q$. Here $M = 0.9, \Lambda = 0.076$ and $ l = 2$}.
\label{omegacharge}
 \end{center}
 \end{figure}

\begin{figure} [H]
\begin{center}
\includegraphics{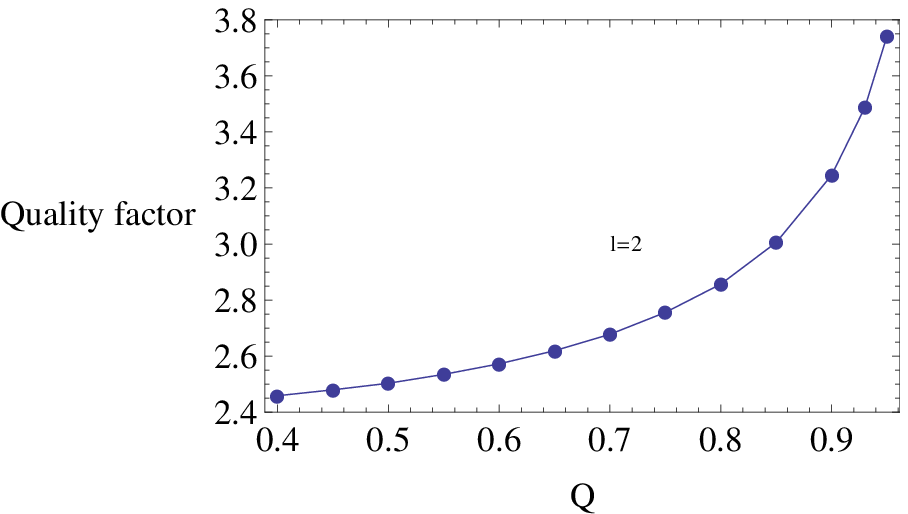}
\caption{The figure shows  Quality factor vs $Q$. Here $M = 0.9, \Lambda = 0.076$ and $ l = 2$}
\label{qualitycharge}
 \end{center}
 \end{figure}
 
 \begin{figure} [H]
\begin{center}
\includegraphics{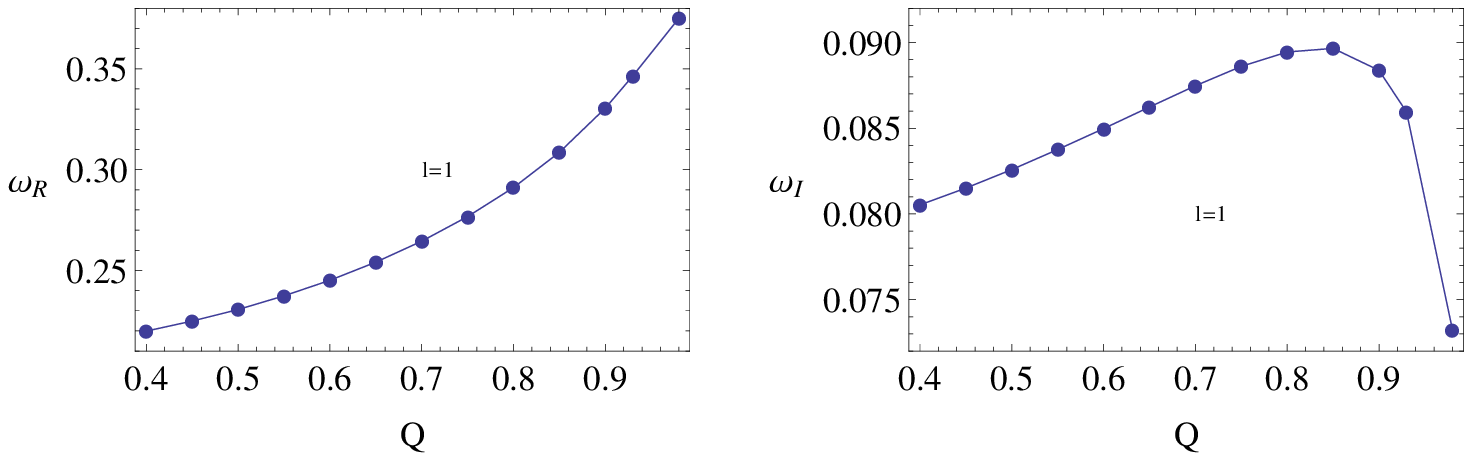}
\caption{The figure shows  $\omega$ vs $Q$. Here $M = 0.9, \Lambda = 0.076$ and $ l = 1$}.
\label{omegacharge2}
 \end{center}
 \end{figure}

\begin{figure} [H]
\begin{center}
\includegraphics{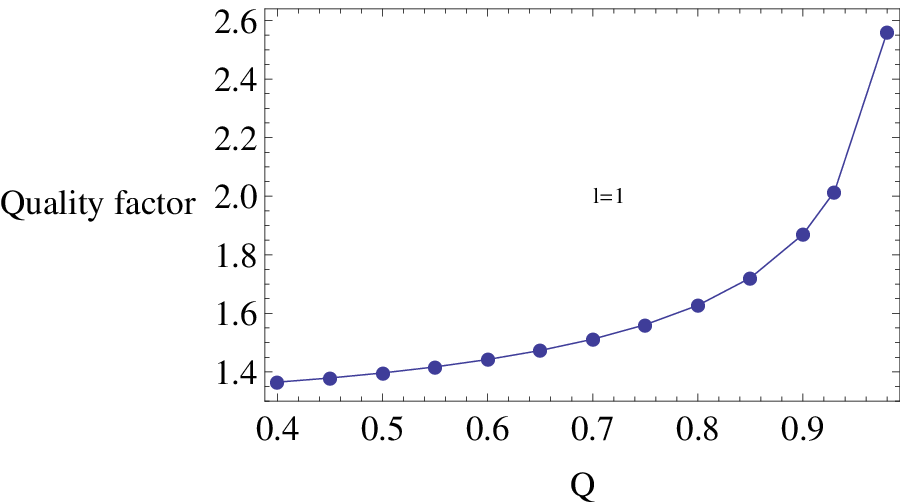}
\caption{The figure shows  Quality factor vs $Q$. Here $M = 0.9, \Lambda = 0.076$ and $ l = 1$}
\label{qualitycharge2}
 \end{center}
 \end{figure}

 \begin{figure} [H]
\begin{center}
\includegraphics{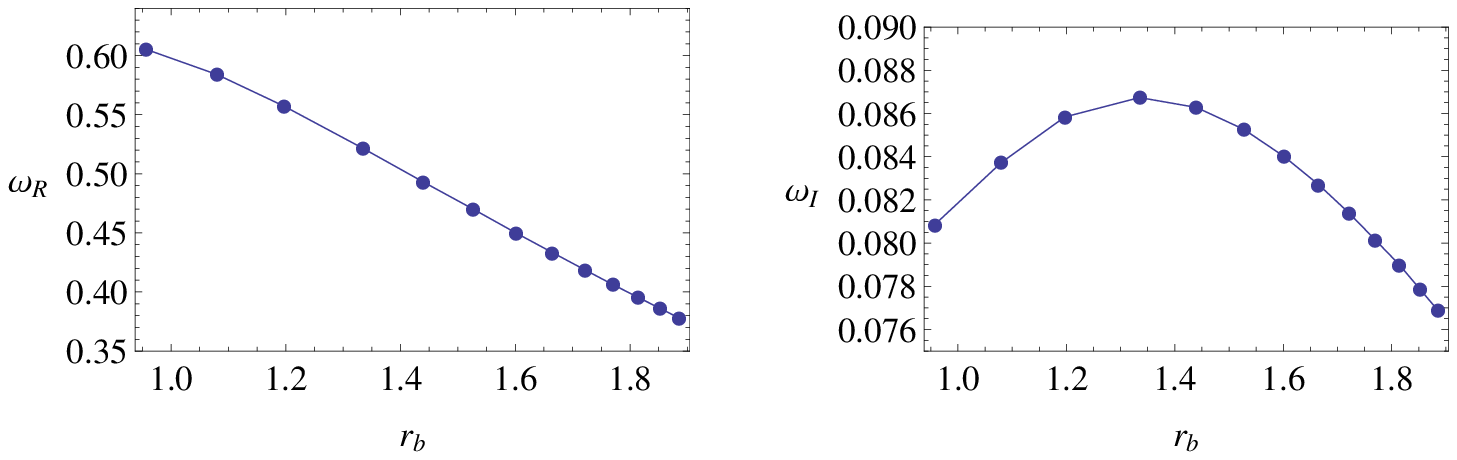}
\caption{The figure shows  $\omega$ vs $r_b$. Here, $M =0.9, \Lambda =0.076$ and $l=2$}
\label{omegaradius}
 \end{center}
 \end{figure}

 \begin{figure} [H]
\begin{center}
\includegraphics{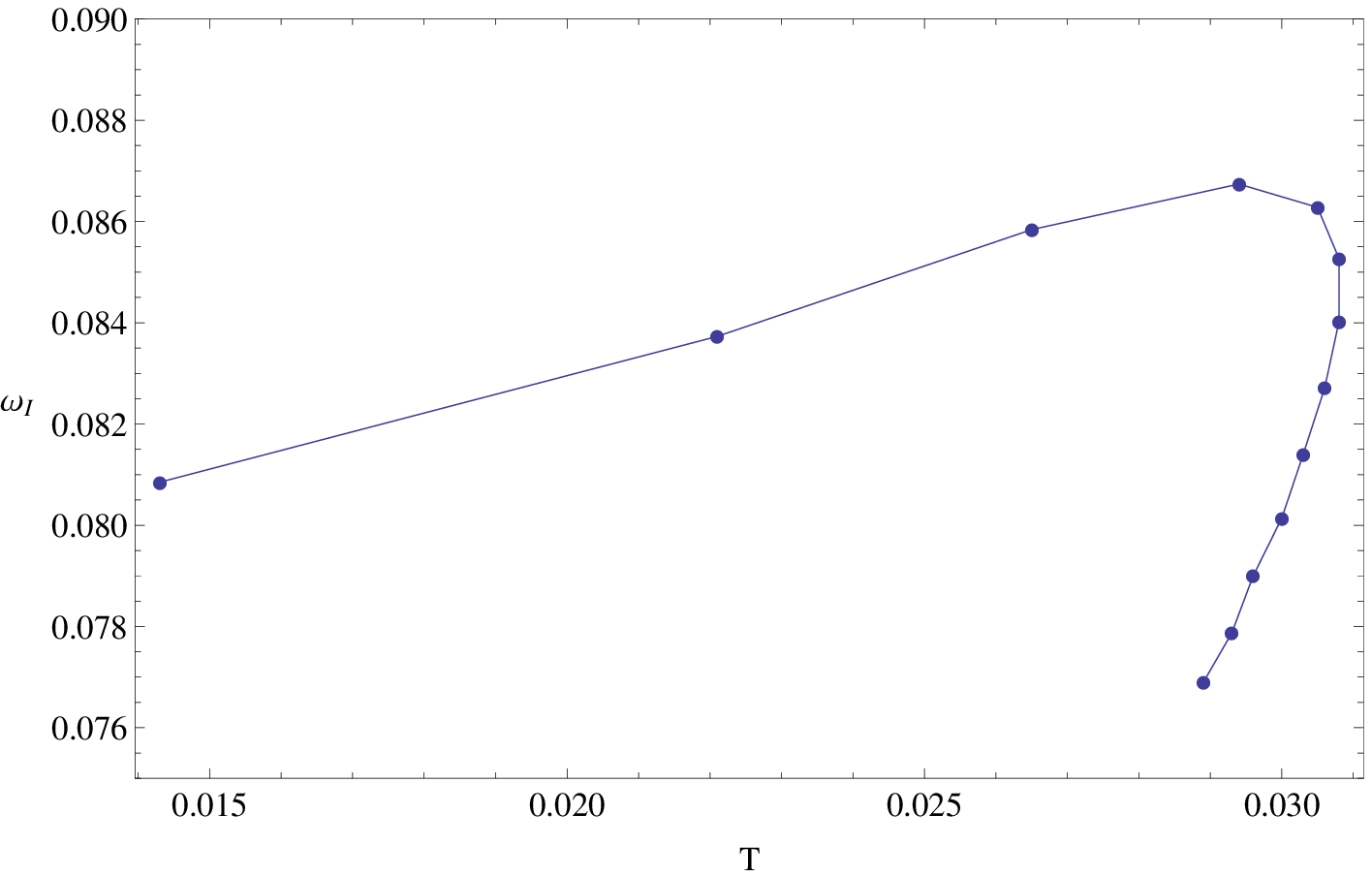}
\caption{The figure shows  $\omega_I$ vs $T_b$. Here, $M =0.9, \Lambda =0.076$ and $l=2$}
\label{omegatemp}
 \end{center}
 \end{figure}

\begin{figure} [H]
\begin{center}
\includegraphics{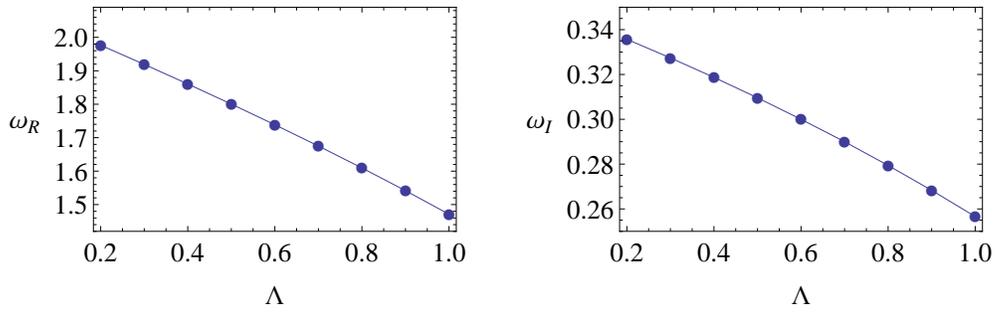}
\caption{The figure shows  $\omega$ vs $ \Lambda$. Here $Q = 0.25, M = 0.278$ and $l=2$}
\label{omegacosmo}
 \end{center}
 \end{figure}

\begin{figure} [H]
\begin{center}
\includegraphics{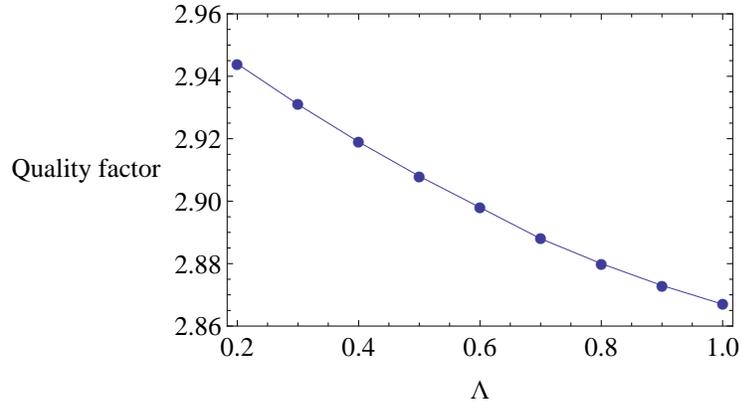}
\caption{The figure shows  Quality factor vs $\Lambda$. Here $Q = 0.25, M = 0.278$ and $l=2$}
\label{qualitycosmo}
 \end{center}
 \end{figure}

\begin{figure} [H]
\begin{center}
\includegraphics{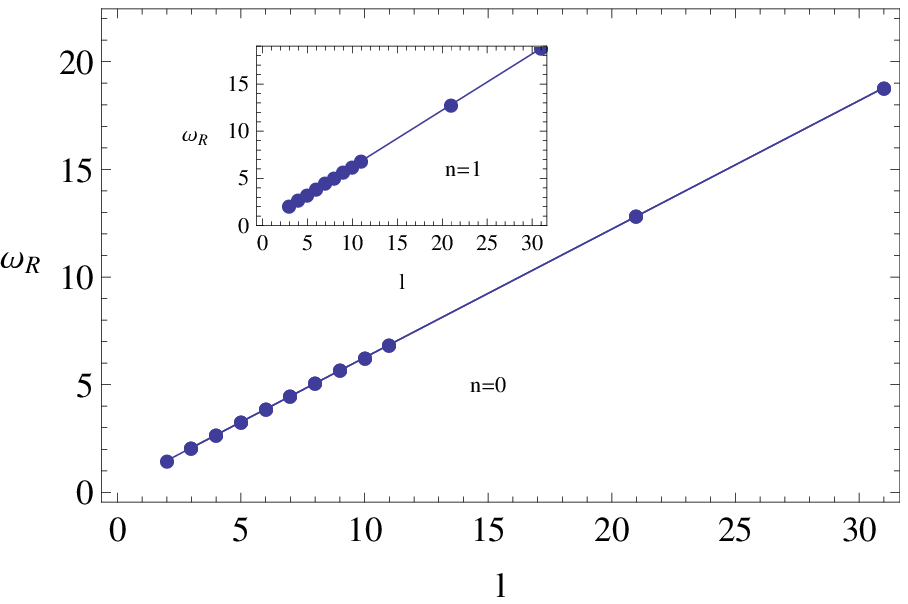}
\caption{The figure shows   $\omega_R$ vs $l$. Here, $Q =0.25, M = 0.278$ and $\Lambda =1$}
\label{sphereal}
 \end{center}
 \end{figure}

\begin{figure} [H]
\begin{center}
\includegraphics{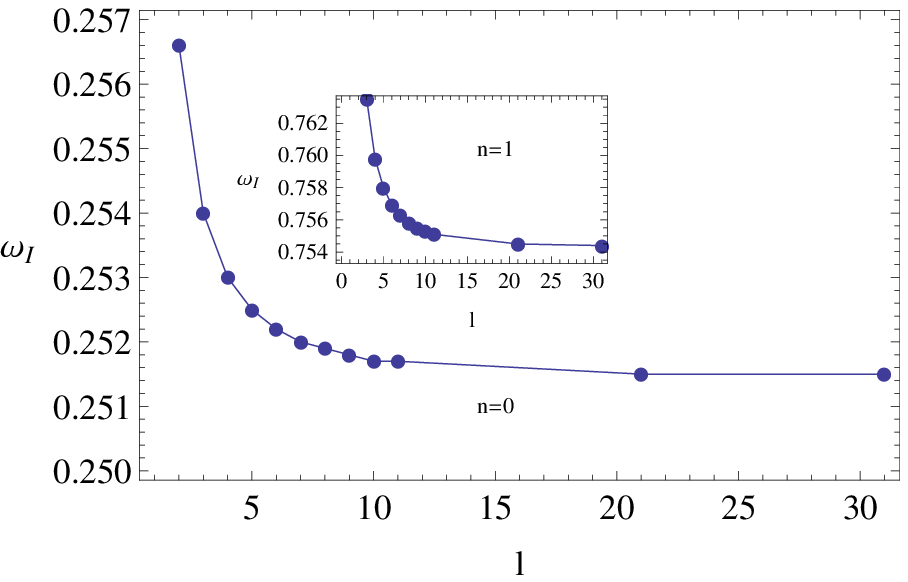}
\caption{The figure shows $\omega_I$ vs $l$. Here, $Q =0.25, M = 0.278$ and $\Lambda =1$}
\label{spheima}
 \end{center}
 \end{figure}

\begin{figure} [H]
\begin{center}
\includegraphics{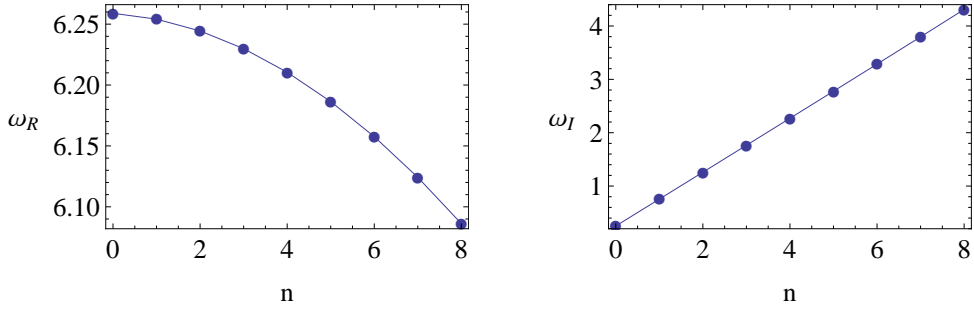}
\caption{The figure shows  $\omega$ vs $n$. Here $Q =0.25, M = 0.278, l=10$ and $\Lambda =1$}
\label{omegan}
 \end{center}
 \end{figure}
 

\section{ Analysis of $l=0$ mode of massless scalar field}

As given in Fig.$\refb{pot2}$, potential for $l=0$ modes is rather different from all other potential with $ l>0$.  The difference is attributed to the fact that the potential is negative in a range between the black hole horizon and  the cosmological horizon. In fact, similar behavior persists for $l=0$ potential for both Schwarzschild-de Sitter black hole and the Riessner-Nordstrom-de Sitter black hole as  shown in Fig$\refb{potboth2}$.

\begin{figure} [H]
\begin{center}
\includegraphics{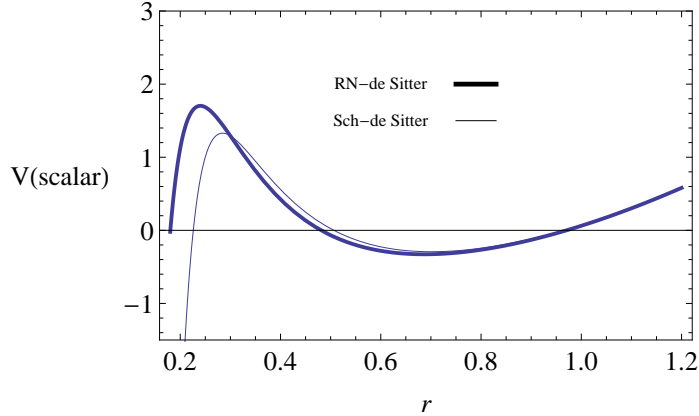}
\caption{$V(r)$ vs $r$ for $l=0$ mode for Reissner-Nordstrom-de Sitter and Schwarzschild-de Sitter black hole. Here, $M=0.1082, \Lambda = 2.5$ and $Q = 0.086$}
\label{potboth2}
 \end{center}
 \end{figure}

WKB approximation that was used in section(5.1) to compute the QNM frequencies of $l>0$ modes are not suitable to compute the  frequencies of $l=0$ modes since there is an extra local minimum of the potential. Hence we will avoid computing the QNM frequencies for this case. Instead, we would be  interested in understanding if the black hole is stable under the $l=0$ mode perturbations. In fact, when the potential is negative in some region, it is possible to have growing QNM modes leading to instability of the system under such perturbations. However, it turns out that this is not the case for all cases with negative potentials as discussed in \cite{kono14}. What we would present here is to the fact that the field of $l=0$ mode in fact settle in to a constant value at late times. Since one of the reasons to study QNM frequencies is to understand the stability of the black hole under such perturbations, such a study is appropriate.  If one needs to compute the QNM frequencies, and see how the field evolves, then a time-domain integration method used by Gundlach et.al. in \cite{gund} can be used. Instead, we will present an analytical observation of how the field behaves at late time.

$l=0$ mode does approach a constant value for the Schwarzschild-de Sitter black hole as presented by Brady et.al \cite{brady}. We will use similar approach to show the behavior of the field around  the regular-de Sitter black hole.

Here, we will study the scattering of the field between $r_b$ and $r_c$ for $l=0$ mode. First we will introduce null coordinates $u$ and $v$ as,
\be
u = t - r_*; \hspace{1 cm} v = t + r_*
\ee
Here, $r_*$ is the tortoise coordinate introduced in eq.$\refb{tor}$. 
Then, the line element in eq.$\refb{metric}$ simplifies to
\be 
ds^2 = - f(r) du dv + r^2 d \Omega^2
\ee
According to the null coordinates, the future cosmological horizon, $r = r_c$ is located at $v=\infty$ and the future black hole horizon, $r_b$ is located at $ u = \infty$. Now, in terms of the null coordinates, the scalar wave equation  $\bigtriangledown^2  \eta = 0$ for a massless scalar field simplifies to,
\be \label{wave3}
\frac{ \partial^2 \Psi(u,v)}{\partial u  \partial v} = - \frac{1}{4} V_{l=0}(r) \Psi(u,v)
\ee
where,
\be
\eta(u,v, \theta, \phi) = \sum_{l,m} Y_{l,m}(\theta, \phi) \frac{ \Psi(u,v)}{r}
\ee
and
\be
V_{l=0} =\frac{ f(r) f'(r)}{r}
\ee
Notice that for $l=0$, the potential simplifies.

When $l=0$, the solution to the eq.$\refb{wave3}$ can be expressed as a series depending on two arbitrary functions $G(u)$ and $H(v)$ as  \cite{brady}\cite{gund},
\be \label{expansion}
\Psi = A \left( G(u) + H(v) \right) +  \sum_{p=0}^{\infty} B_p(r) [ G^{(-p-1)} (u) + (-1)^{p+1} H(v)^{(-p-1)} ]
\ee
Here, the negative super indices on $G$ and $H$ refer to integration with respect to $u$ and $v$ respectively: for example, $G^{-1}(u) = \int G(u) du$. The coefficient $A$ can be set equal to $1$ without loss of generality.

When eq.$\refb{expansion}$ is substituted to eq.$\refb{wave3}$, one can compare the coefficients to come up with a recurrence relations for $B_p(r)$ as,
\be \label{recur}
B_{p+1}' = \frac{f'}{ 2 r} \left( - B_p + r B_p' \right)  +  \frac{ f}{2} B_p''
\ee
Here,
\be \label{bnote}
B_0 = - \int \frac{f'}{2 r} dr
\ee
To facilitate the computation, we will assume the charge $Q$ is small and do an expansion for the function $f(r)$ for  the regular de-Sitter black hole as,
\be
f(r) \approx 1 - \frac{ 2 M}{r} + \frac{ Q^2}{ r^2} - \frac{ Q^6}{ 12 M^2 r^4}
\ee
Note that the qualitative behavior of the black hole outside $r_b$ and inside $r_c$ are not much different in the expansion. Now, one can integrate eq.$\refb{bnote}$ to obtain $B_0$ as,
\be
B_0 = \frac{ Q^6}{ 30 M^2 r^5} - \frac{ Q^2}{ 3 r^3} + \frac{M}{ 2 r^2} + \frac{ r \Lambda} { 3}
\ee
From the recurrence relation in eq.$\refb{recur}$, $B_1$ can be obtained to be,
\be
B_1 = \sum_{ n=3}^{ 11} \frac{ a_n}{r^n}
\ee
where $a_n$ depends on $M, Q$ and $\Lambda$.  Other $B_p$ values with $ p>1$ also behaves similarly with higher powers of  $\frac{1}{r}$ terms. For large $r$, $B_p$ will vanish except for $B_0$. This is one of the major difference compared to an asymptotically flat black hole where all $B_p$ values will vanish for large $r$.

Now, we consider an initial burst of radiation on $v =0$. The radiation is confined between $ u=0$ and $ u = u_1$ and is given by $\Psi(u,0) = G(u)$. Also, $\Psi(0,v) =0$. This is demonstrated in Fig.$\refb{draw}$. We want to study the evolution of this field in the diamond shape region given in Fig.$\refb{draw}$  by thick lines.

\begin{figure} [H]
\begin{center}
\includegraphics{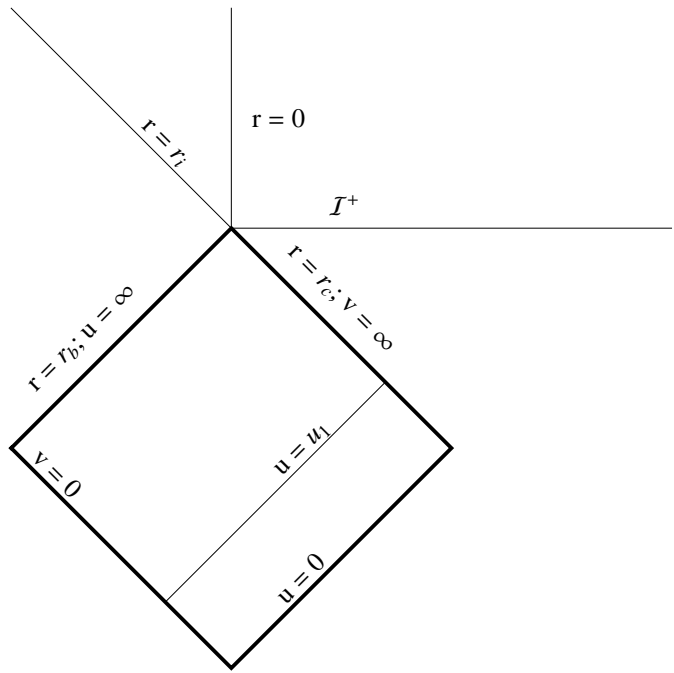}
\caption{The region in space-time outside the black hole studied for the evolution of an initial burst confined between $u=0$ and $u=u_1$ and on $v=0$. The initial bust is such that, $\Psi(u,0) = G(u)$ where $G(u) \neq 0$, for $0<u<u_1$ only.}
\label{draw}
 \end{center}
 \end{figure}

Following the work by Gundlach et.al \cite{gund} and Brady et.al. \cite{brady}, the evolution of the initial burst can be broken into two parts:
\noindent
(i) the evolution of the burst $\Psi(u, 0) = G(u)$ which is non-zero between $0<u<u_1$ and $\Psi(0,v) = 0$.
(ii) subsequent evolution in the region for $u \geq u_1$.


\subsection{ The evolution of the burst $\Psi(u, 0) = G(u)$ which is non-zero between $0<u<u_1$ and $\Psi(0,v) = 0$ }

Since the initial burst is independent of $v$ and $G(u_1) =0$, the field at $u = u_1$ is given by,
\be \label{A4}
\Psi(u_1, v) = \sum_{p=0}^{\infty} B_p(r) G(u)^{(-p-1)} (u_1)
\ee
For sufficiently small $\Lambda$, $r_c$ is large and at $ r= r_c$, the only non-zero $B$ is 
$B_0 \approx  \frac{ \Lambda r_c}{3}$. Hence, at the cosmological horizon $r_c$ ($ v = \infty$),

\be \label{A5}
\Psi(u_1, \infty) \approx  B_0 G^{-1}(u)  =  \frac{ \Lambda r_c}{ 3} \int^{u_1}_{0}  G(u) du + O(\Lambda)
\ee

\subsection{ The evolution of the field in $ u  \geq u_1$ region}

Now, we can examine the evolution of the field in the region $ u  \geq u_1$ by taking eq.$\refb{A4}$ as the initial data.  Following the work in \cite{brady} and \cite{gund}, the field equation given in eq.$\refb{wave3}$ can be solved near the cosmological horizon as,
\be \label{A52}
\Psi(u,v) \approx [ G(u) + H(v) ]  + B_0(r) [ G^{-1}(u) - H^{-1}(v) ]
\ee
Defining $G^{-1}(u) = g(u)$ and $H^{-1}(v) = -h(v)$ where $g(u)$ and $h(v)$ are arbitrary functions. Approximating $B_0$ to be $\frac{ \Lambda r}{3}$ since $r$ is large, one can rewrite eq.$\refb{A6}$ as,
\be \label{A6}
\Psi(u,v) \approx \frac{ \Lambda  r}{ 3} [ g(u) + h(v) ] + [ \frac{ \partial g(u)}{ \partial u} - \frac{ \partial h(v)}{ \partial v} ]
\ee
Since closer to the cosmological horizon $r_c$, $r_* \approx -\frac{1}{f'(r_c)} Log( r_c - r)$ from eq.$\refb{torcos}$, and, also the fact that $ r_*= \frac{ v- u}{2} $, one can write $r$ as,
\be
r \approx r_c - e^{  -\kappa_c v} e^{ \kappa_c u}
\ee
Here $\kappa_c = \frac{1}{2} \left|\frac{ df'(r)}{dr}\right|_{r=r_c}$ is the surface gravity at $r_c$. Hence closer to the cosmological horizon (when $ v \ra \infty$), the right hand side of eq.$\refb{A4}$ can be approximated with a polynomial of $e^{- \kappa_c v}$. Hence by comparing eq.$\refb{A4}$ and eq.$\refb{A6}$, one can conclude that the function $h(v)$ is also a polynomial of $e^{- \kappa_c v}$ given as,
\be \label{A7}
h(v) = \sum_{n=0}^{\infty} h_n e^{- n \kappa_c v}
\ee
When $ v \ra \infty$, one can compute $h_0$ from eq.$\refb{A5}$ and eq.$\refb{A6}$,  as,
\be
h_0 = \frac{ 3}{ \Lambda r_c} \Psi(u_1, \infty)
\ee
which is non-zero.
The nature of the function $g(u)$ depends on the potential everywhere. However, since $g(u)$ also evolves from eq$\refb{A4}$, it seems reasonable to predict that it is of the form (when $ u \ra \infty$)
\be
g(u) = \sum_{n=0}^{\infty} g_n e^{- n \kappa_c u}
\ee
Hence, at late times, the  field $\Psi \ra h_0 + g_0$ and the scalar field $\eta =  \frac{ \Psi}{r} = \frac{ h_0 + g_0}{r_{c,b}}$ approach a constant value  at late times (when $u$ and $v$ both approach $\infty$). This is only possible only when $h_0 + g_0 \neq 0$. When $l=0$ mode, this is the only static solution which is regular at both horizons $r_b$ and $r_c$. Hence, the field decay into a constant value late times only for this particular mode without decaying completely. Therefore, $l=0$ mode of the massless scalar field is prone for instability.


\section{ Quasi normal modes of massless fields from  null geodesics}

The black holes considered in this paper have unstable null geodesics. Cardoso et.al. \cite{cardoso1}, showed that in the eikonal limit ($l>>1$), QNM of a black hole can be determined by the parameters of null geodesics. An earlier work done along those lines is given by   Mashhon \cite{mas1} 

An expansion method to compute QNM based on the link between the null geodesics and QNM freuencies has been applied to Kerr black hole   in a paper by Dolan \cite{dolan1}.  The method of null geodesics  has been applied to study the near-extreme Kerr black hole by Hod \cite{hod1} and black holes in anti-de Sitter space by Morgan et.al in  \cite{morgan}.

The equation of motion for massless particles around a black hole is given by \cite{fernando12}

\begin{equation}
\dot{r}^2 + V_{null}= E^2
\end{equation}
with,
\begin{equation}
V_{null} = \left(  \frac{L^2}{r^2}   \right) f(r) 
\end{equation}
Here, $L$ is the angular momentum of the massless particle. For bot $ r = r_h$ and $ r = r_c$, $ V_{null} = 0$. In the Fig.$\refb{potnull}$,   $V_{null}$ is plotted for various values of the magnetic charge $Q$.  One can observe that the height of the potential  is higher for higher values of  $Q$.

\begin{figure} [H]
\begin{center}
\includegraphics{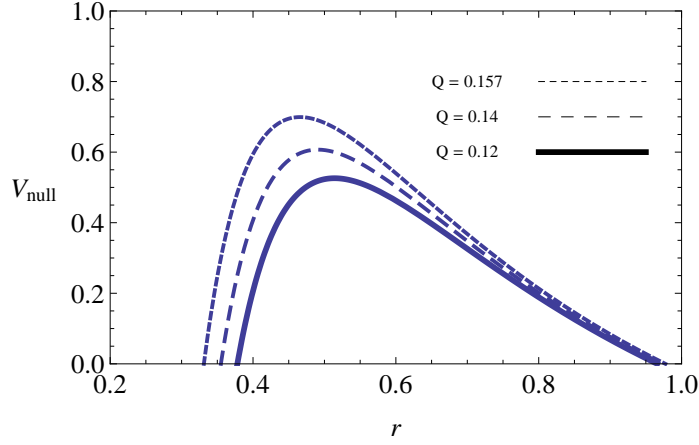}
\caption{The figure shows  $V_{null}(r)$ vs $r$. Here $M = 0.19, L=1$ and $\Lambda =2$}
\label{potnull}
 \end{center}
 \end{figure}
 
Now, when   the effective potential for the massless scalar field $V_{scalar}$ in eq.(17) is studied, one can see that in the eikonal limit ($ l \rightarrow \infty$),
\begin{equation}
V_{scalar}  \approx  \frac{ f(r) l}{r^2}
\end{equation}
Hence,  one can conclude that the maximum of $V_{scalar}$ occurs at $r = r_{max}$ given by,
\begin{equation}
2 f(r_{max})  - r_{max} f'(r_{max}) =0
\end{equation}
On the other hand, the maximum of $V_{null}$  occurs at $V_{null}'=0$ leading to,
\begin{equation}
2 f(r_{null})  - r_{null} f'(r_{null}) =0
\end{equation}
Hence the maximum of $V_{scalar}$ and the location of the maximum of the null geodesics coincides at $ r_{max} = r_{null}$.  $V_{null}(r)'=0$ leads to the equation,
\be
\frac{ Q^2}{r^5} Sech^2\left( \frac{ Q^2}{ 2 M r}\right) + \frac{ 2}{ r^4} \left( - 3 M + r + 3 M  Tanh\left( \frac{ Q^2}{ 2 Mr}\right) \right) = 0
\ee
Solutions to the above equation gives $r_{null}$ values. Since an analytical expression is not possible, we have calculated $r_{null}$ varying charge $Q$ numerically and plotted in Fig.$\refb{rnull}$. One can observe that $r_{null}$ gets smaller as $Q$ increases.

\begin{figure} [H]
\begin{center}
\includegraphics{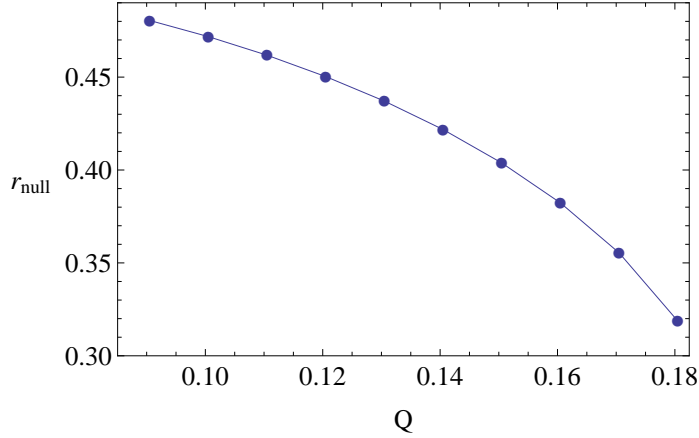}
\caption{The figure shows  $r_{null}$ vs $Q$. Here $M = 0.1715, L=1$ and $\Lambda =0.0973$}
\label{rnull}
 \end{center}
 \end{figure}
 
According to \cite{cardoso1}  the QNM frequencies in the eikonal limit is given by,
\begin{equation} \label{nullfre}
\omega_{QNM} = \Omega_a l - i ( n + \frac{1}{2} ) | \lambda_{Ly}|
\end{equation}
Here, $l$ is the angular momentum of the perturbation and $n$ is the overtone number, and $\Omega_a$ is the coordinate angular velocity given as,
\begin{equation} \label{omeganull}
\Omega_a=  \frac{\dot{\phi}}{\dot{t}}
\end{equation}
The parameter  $\lambda_{Ly}$ appearing in eq.$\refb{nullfre}$ is the Lyapunov exponent which is interpreted as  the decay rate  of the unstable circular null geodesics. The derivation of the above results  is explained in Cardoso et.al\cite{cardoso1}. For the ABGB-dS black hole, $\Omega_a$ and $\lambda_{Ly}$ are given as,
\begin{equation}
\Omega_a = \frac{ \dot{\phi}(r_{null})}{\dot{t}(r_{null})}  = \sqrt{ \frac{ f(r_{null})}{r_{null}^2 }} 
\end{equation}

\begin{equation} \label{lambdanull}
\lambda_{Ly} = \sqrt{ \frac{ -V_{null}''(r_{null})}{ 2 \dot{t}(r_{null})^2}} 
= \sqrt{ \frac{-V_{null}''(r_{null}) r_{null}^2 f(r_{null})}{2 L^2} }
\end{equation}

By using $\Omega_a$ and $\lambda_{Ly}$, the QNM frequencies are calculated and plotted in Fig.$\refb{lya}$ and Fig.$\refb{angular}$. When compared with $\omega$ values computed with WKB approach given in Fig$\refb{omegacharge}$, one can conclude that they are similar in behavior.

\begin{figure} [H]
\begin{center}
\includegraphics{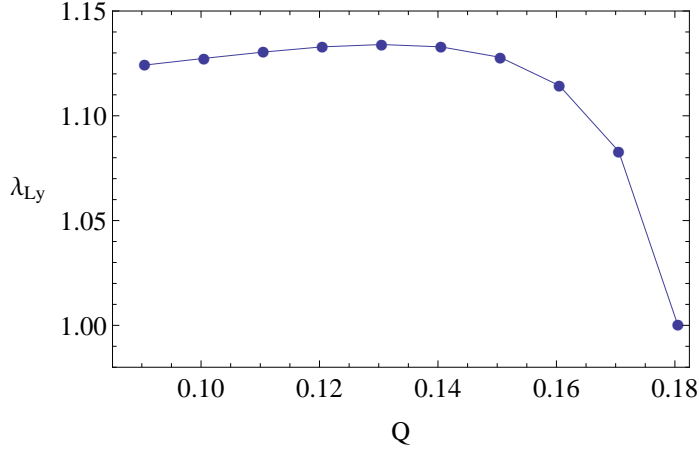}
\caption{The figure shows  $\lambda_{Ly}$ vs $Q$. Here $M = 0.1715, L=1$ and $\Lambda =0.0973$}
\label{lya}
 \end{center}
 \end{figure}

\begin{figure} [H]
\begin{center}
\includegraphics{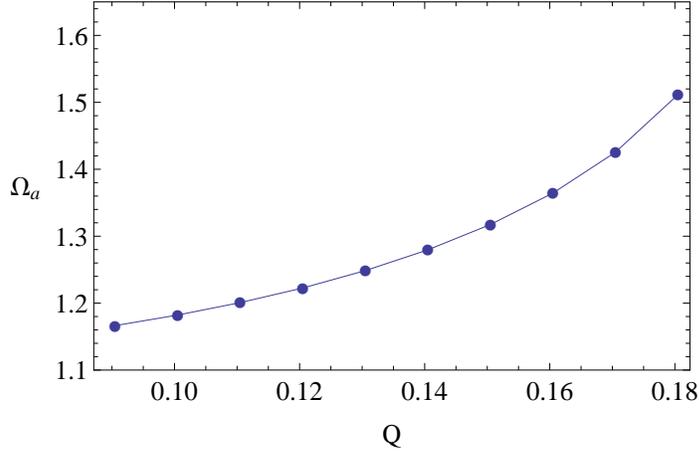}
\caption{The figure shows  $\Omega_a$ vs $Q$. Here $M = 0.1715, L=1$ and $\Lambda =0.0973$}
\label{angular}
 \end{center}
 \end{figure}



\section{ P$\ddot{o}$schl-Teller approximation for the near-extreme  ABGB-de Sitter black hole}

First we like to discuss the extreme ABGB-dS black hole  when $ r_b = r_c$.  For the black hole to  be extreme, two conditions has to be satisfied:
\be 
f(r) =0;  \hspace{ 1 cm}  f'(r) =0
\ee
Due to the complex nature of the function $f(r)$, an exact value for $r_b$ for the extreme black hole is not possible. Let us name the value of $r_b$ at this special case as $\sigma$. It is clear that $f''(\sigma) <0$ due to the nature of the function $f(r)$. Such extreme black holes where $r_b = r_c$ are called Nariai black holes \cite{fernando3}\cite{fernando4}\cite{fernando5}.

In the near extreme case, $r_b$ and $r_c$ are very close. Hence $f(r)$ can be expanded in a Taylor series as \cite{mat2} as,

\begin{equation}
f(r) \approx  \frac{ f''(\sigma)}{2} ( r - r_b) ( r - r_c)
\end{equation}
Hence the tortoise coordinate defined  as $r_* = \int \frac{dr}{f(r)}$ can be integrated as,
\be \label{newtr}
r_* = \int \frac{dr}{ f(r)} =   \frac{2}{ f''(\sigma) ( r_c - r_b)} log\left( \frac{ r_c - r}{ r - r_b} \right) 
\ee
From the above derivation, it is clear that when $r \ra r_b$, $r_* \ra -\infty$ and $ r \ra r_c$, $r_* \ra \infty$. 
\noindent
From the relation in eq.$\refb{newtr}$ can be solved for $r$ as,
\be \label{rvalue}
r = \frac{ r_b + r_c e^{ \rho r_*}}{ 1 + e^{ \rho r_*}}
\ee
Here,
\be \label{eta}
\rho = -\frac{f''(\sigma)}{2}  ( r_c - r_b) 
\ee
Bu substituting $r$ in   eq.$\refb{rvalue}$, the function $f(r)$ for the near extreme ABGB-de Sitter black hole can be written as,
\be \label{newfr}
f(r)  =  \frac{ \xi}{ 4( Cosh( \frac{ \rho r_*}{2}))^2} 
\ee
Here,
\be
\xi =  -\frac{ f''(\sigma)}{2} ( r_c - r_b)^2
\ee
Now the effective potential for the scalar field can be written as,
\be
V_{eff} =  \frac{ V_0}{ Cosh^2( \frac{\rho r_*}{2})}
\ee
where
\be \label{v0}
V_0 = \frac{ \xi l ( l+1)}{ 4 \sigma^2}  + \frac{\xi}{4} \mu^2
\ee
In deriving $V_0$,  we have assumed  $ r  \approx \sigma$. Also since $f(r)' \approx 0$, only the first term in the potential dominates. With the new definition of the potential,  the wave equation for the scalar field perturbations simplifies to,
\be \label{wave2}
\frac{ d^2 \Omega(r_*) }{ dr_*^2} + \left( \omega^2 -   \frac{ V_0}{ Cosh^2( \frac{\rho r_*}{2})}  \right) \Omega(r_*) = 0
\ee
$\frac{ V_0}{ Cosh^2( \frac{\rho r_*}{2})}$ is the well known   P$\ddot{o}$schl-Teller potential and Ferrari and Mashhoon,  in a well known paper \cite{ferra3} demonstrated that  $\omega$ can be computed exactly as,
\be
\omega =  \sqrt{ V_0  - \frac{\rho^2}{16}} - i \frac{\rho}{2} ( n + \frac{1}{2})
\ee
By observing the expression for $\omega$ given above, some hand waving arguments of the behavior of $\omega$ can be made.  When $\mu$ is large,  and $l$ is small,
$V_0 \approx \frac{ \xi \mu^2}{ 4}$. Hence, $\omega_R \approx \frac{ \sqrt{\xi} \mu}{ 2} + constant$ and $\omega_R$ will depend linearly on $\mu$. On the other hand, $\omega_I$ does not depend on $\mu$  for extreme black holes. When the spherical harmonic index $l$ is large, $V_0 \approx \frac{ \xi l^2}{ 4 \sigma^2}$. Hence $\omega_R \approx \frac{ \sqrt{\xi} l} { 2 \sigma} + constant$. Hence $\omega_R$ depends on $l$ linearly  for large $l$ which is clear from Fig$\refb{sphereal}$. For large $l$, $\omega_I$ is independent of $l$ and is clear from Fig$\refb{spheima}$. For large $n$, $\omega_I$ becomes large and $\omega_R$ is independent of $n$. This behavior is represented in Fig.$\refb{omegan}$.


\section{ Conclusion}

Our main goal in this paper has been to study QNM and stability of a regular black hole in de Sitter universe under scalar perturbations. The regular black hole does not have a singularity at $ r =0$. It has three horizons: cosmological horizon $r_c$, black hole event horizon $r_b$ and inner horizon $r_i$. The structure of the regular-de Sitter black hole is very similar to the Reissner-Nordstrom-de Sitter black hole.

We have employed sixth order WKB approximation to compute QNM frequencies. The parameters of the theory, the mass $M$, magnetic charge $Q$, cosmological constant $\Lambda$ and the spherical index $l$ and the mode number $n$ were changed to see how QNM frequencies depend on them. When $Q$ is  increased, $\omega_R$ increases. On the other hand, when $Q$ is increased, $\omega_I$ increases to a maximum and then decreases for large $Q$. Hence, there is a particular value of $Q$ where the  black hole have the most stability. We have observed the behavior of the quality factor for the black hole as a function of the charge $Q$. It is noticed that the larger the charge,  larger the quality factor is.

Next, we have plotted $\omega$ as a factor of the black hole $r_b$. When $r_b$ increases, $\omega_R$ decreases and $\omega_I$ increases to a maximum and then decreases. Hence there is a particular size of the black hole where the stability is maximized. We have also computed the Hawking temperature to observe the relation with $\omega_I$.  The relation is not linear as in other black holes. 

We have studied the behavior of $\omega$ with the cosmological constant $\Lambda$. When $\Lambda$ is increased,  both $\omega_R$ and $\omega_I$ decreases. Hence, a smaller $\Lambda$ is favored for a stable black hole. The behavior of $\omega$ with respect to $l$, the spherical harmonic index is similar to the behavior of other black holes. $\omega_R$ has a linear relation with $l$ and $\omega_I$ is independent of $l$ for large $l$. We have also plotted  $\omega$ vs $n$, the mode number. Here, $\omega_R$ decreases with $n$ and  $\omega_I$  increases linearly with $n$. 

The $l=0$ mode for the massless field presents a potential with a negative region between $r_b$ and $r_c$. WKB approach cannot be used to compute QNM frequencies in such a case since there is an additional local minimum in the potential. Instead, we did an analytical study to show that an initial burst of radiation will settle into a constant value at late times. Such behavior have been observed for the $l=0$ mode for the d Schwarzschild-de Sitter black holes.

In addition to the WKB approach, we have used the unstable null geodesics approach to compute $\omega$ for massless scalar fields.  First we observe that the radius of the unstable null geodesic $r_{null}$ decreases with $Q$. The coordinate angular velocity, which is proportional to $\omega_R$ increases with $Q$. The Lyapunov exponent which is proportional to $\omega_I$  does  show similar behavior as was done using WKB approach.

Finally, we have demonstrated that for the near-extremal black hole with $r_b \approx r_c$ the scalar field equation will have the P$\ddot{o}$schl-Teller potential; one can get exact expressions  for he QNM frequencies for this observation. Most of the observation we did with WKB approach,  tally with the exact expressions obtained via this method.

As future work, it may be interesting to compute QNM frequencies and study stability of the massive fields. We did not use WKB approach to compute QNM frequencies of this case since there may be extra  local minima for the potential. One may use continued fraction method used by Konoplya and Zhidenko  \cite{kono5} to compute massive QNM for the Schwarzschild black hole.

\vspace{0.5 cm}


{\bf Acknowledgments:}  The author wish to thank R. A. Konoplya for providing the  {\it Mathematica}  file for  the WKB approximation.


\end{document}